\documentclass[12pt]{article}

\usepackage{comment}
\usepackage{arxiv}
\usepackage{graphics}
\usepackage{graphicx}
\usepackage{makecell}
\usepackage{hyperref}
\usepackage{amsmath}
\usepackage{IEEEtrantools}
\usepackage{relsize}%used for \mathlarger{} command
\usepackage[UKenglish]{babel}
\usepackage{csquotes}
\usepackage[sorting=none]{biblatex} 
\usepackage{parskip}
\usepackage{placeins}
\usepackage{caption} 
\usepackage{titlesec}
\titlespacing*{\section}{0pt}{1.7\baselineskip}{1.05\baselineskip}
\usepackage{setspace}
\nocite{*}

\title{The Principle of equal Probabilities of Quantum States}

\author{ \href{https://orcid.org/0000-0001-6088-8919}{\includegraphics[scale=0.06]{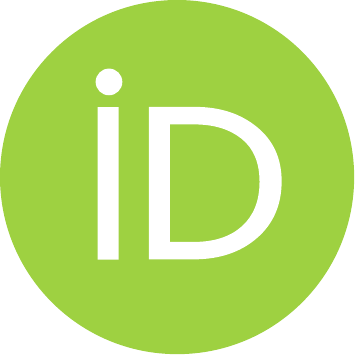}\hspace{1mm}Michalis Psimopoulos}\\
    on leave from Imperial College London\\
	Kanari 26, Pefki, 15121, Athens, Greece\\
	\texttt{m.psimopoulos@hotmail.com} \\
	\And
	\href{https://orcid.org/0000-0003-2496-8393}{\includegraphics[scale=0.06]{orcid.pdf}\hspace{1mm}Emilie Dafflon} \\
	Athinon 109, Voula, 16673, Athens, Greece\\
	\texttt{emsdafflon@gmail.com} \\
}

\hypersetup{
pdftitle={The Principle of equal Probabilities of Quantum States},
pdfauthor={Michalis Psimopoulos, Emilie Dafflon},
}

\begin{document}

%\sloppy

\maketitle

%\sloppy
\begin{abstract}
%\noindent
The statistical problem of the distribution of $s$ quanta of equal energy $\epsilon_0$ and total energy $E$ among $N$ \textit{distinguishable} particles is resolved using the conventional theory based on Boltzmann's principle of \textit{equal probabilities of configurations} of particles distributed among energy levels and the concept of \textit{average state}. In particular, the probability that a particle is in the $\kappa$-th energy level i.e. contains $\kappa$ quanta, is given by
\begin{equation*}
  p(\kappa)=\displaystyle \frac{\displaystyle \binom{N+s-\kappa-2}{N-2}}{\displaystyle \binom{N+s-1}{N-1}} \;\;\; ; \;\;\; \kappa = 0, 1, 2, \cdots, s
\end{equation*}
In this context, the special case ($N=4$, $s=4$) presented indicates that the alternative concept of \textit{most probable state} is not valid for finite values of $s$ and $N$.\newline
In the present article we derive alternatively $p(\kappa)$ by distributing $s$ quanta over $N$ particles and by introducing a new principle of \textit{equal probability of quantum states}, where the quanta are \textit{indistinguishable} in agreement with the Bose statistics.
%/we propose an alternative approach that derives $p(\kappa)$
% the present article we construct a bridge between classical and quantum particle statistics by considering from two points of view the distribution of $s$ quanta of equal energy $\epsilon_{o}$ and total energy $E$ among $N$ \textit{distinguishable} particles.
%\textbf{I.} Using Boltzmann's principle of \textit{equal probabilities of configurations} and the concept of \textit{average state}, it is shown that the probability that a particle is in the $\kappa$-th energy level i.e. contains $\kappa$ quanta, is given by
%\begin{equation*}
%  p(\kappa)=\displaystyle \frac{\displaystyle \binom{N+s-\kappa-2}{N-2}}{\displaystyle \binom{N+s-1}{N-1}} \;\;\; ; \;\;\; \kappa = 0, 1, 2, \cdots, s
%\end{equation*}
%It becomes clear from a special case ($N=4$, $s=4$) that the alternative concept of \textit{most probable state} is not valid for finite values of $s$ and $N$.
%\textbf{II.} Equivalently, it is found that the \textit{same} probability $p(\kappa)$ can also be obtained by distributing $s$ quanta over $N$ particles according to a principle of \textit{equal probabilities of states}, where the quanta are considered to be \textit{indistinguishable} and the Bose statistics is manifested.\newline
Therefore, the analysis of the two approaches presented in this paper highlights the equivalence of quantum theory with classical statistical mechanics for the present system.\newline
At the limit $\epsilon_{o} \rightarrow 0 $; $s \rightarrow \infty $;  $s \epsilon_{o} = E \sim$ fixed, where the energy of the particles becomes continuous, $p(\kappa)$ transforms to the Boltzmann law
\begin{equation*}
  P(\epsilon) = \displaystyle \frac{1}{\langle \epsilon \rangle}e^{-\mathlarger{\frac{\epsilon}{\langle \epsilon \rangle}}} \;\;\; ; \;\;\; 0\leq \epsilon < +\infty
\end{equation*}
where $\langle \epsilon \rangle = E/N$. Hence, the classical \textit{principle of equal a priori probabilities} for the energy of the particles leading to the above law, is justified here by quantum mechanics. %for $N \rightarrow \infty$; $E \rightarrow \infty$; $E/N =  \langle \epsilon \rangle \sim$ fixed.

\end{abstract}
                                    
%\clearpage
%\newgeometry{top=1.0in,bottom=1.5in}
\pagenumbering{gobble}
\clearpage
\pagenumbering{arabic}
\pagestyle{plain}
\setlength\parindent{0pt}
\setlength{\parskip}{2mm}
\section{Introduction}\label{intro}

Consider a system of $N$ \textit{distinguishable} particles having total energy $E$. If the energy is quantised in equal parts: $E=s \epsilon_{o}$, where $s$ is the total number of quanta, the particles will occupy the discrete energy levels $0, \epsilon_{o}, 2\epsilon_{o}, \cdots, s\epsilon_{o}$ (briefly $0, 1, 2, \cdots, s$). In the present article we consider the Boltzmann principle \cite{Boltzmann} of \textit{equal probabilities of configurations} (complexions) for particles distributed among energy levels according to the equations:\begin{align}
\label{eq_particles_N_conserv} & n_0 + n_1 + n_2 + \cdots + n_s = N\\
\label{eq_energy_conserv} & n_1 + 2 n_2 + 3 n_3 + \cdots + s n_s = s
\end{align}
where $n_{0}\geq 0,n_{1} \geq 0, n_{2} \geq 0, \cdots, n_{s} \geq 0$ are the respective number of particles occupying the $s+1$ energy levels. 
As it is well known, each group $(n_{0}, n_{1}, n_{2}, \cdots, n_{s})$ which is a solution of Eqs (\ref{eq_particles_N_conserv}, \ref{eq_energy_conserv}) defines in this case a \textit{state} of the system. 
The total number of states $S_{I}$ created by Eqs (\ref{eq_particles_N_conserv}, \ref{eq_energy_conserv}) can be obtained considering the $s$ dimensional space $[n_{1}, n_{2}, \cdots, n_{s}]$. In this space Eq.(\ref{eq_particles_N_conserv}) defines a family of hyperplanes $n_1 + n_2 + \cdots + n_s = N-n_0$ where the number $n_0$ of particles having zero energy is a parameter $n_0=0, 1, 2, \cdots, N$ and where each hyperplane of this family cuts symmetrically all axes of the $s$ dimensional space at $N - n_0$. The total number of non-negative integer solutions of Eq.(\ref{eq_particles_N_conserv}) is given by the Bose formula \cite{Ladau}:
\begin{equation}\label{Bose_formula}
    {B}(N, s+1) = \sum_{n_{0}=0}^{N} \frac{(N - n_{0} +s -1)!}{(N-n_{0})!(s-1)!} = \begin{pmatrix} N+s \\ s \end{pmatrix}
\end{equation}
On the other hand, Eq.(\ref{eq_energy_conserv}) defines in the $s$ dimensional space $[n_{1}, n_{2}, \cdots, n_{s}]$ a single nonsymmetric hyperplane cutting the axis $n_1$ at $n_{1}=s$, the axis $n_2$ at $n_{2}=\frac{s}{2}$, $\cdots$, the axis $n_s$ at $n_{s}=1$. In this case the number of non-negative integer solutions of Eq.(\ref{eq_energy_conserv}) is equal to the number $p_s$ of partitions of the number $s$. %According to reference \cite{Psimopoulos}:
According to previous work (M. Psimopoulos, ‘Reduced Harmonic Representation of Partitions’, 2011):
\begin{equation}\label{partition_ps_formula}
    p_{s}=\frac{2}{\pi}  \int_{0}^{\frac{\pi}{2}} \prod_{\kappa=1}^{s} \frac{sin((s+\kappa)x)}{sin(\kappa x)} cos[({s^2}-2s)x] \; dx
\end{equation}
Considering next both Eqs (\ref{eq_particles_N_conserv}, \ref{eq_energy_conserv}) we have two cases: \newline
If $N \geq s$ the family of hyperplanes defined by Eq.(\ref{eq_particles_N_conserv}) fully covers the energy hyperplane defined by Eq.(\ref{eq_energy_conserv}) so that the total number of states ${S}_{I}$ corresponding to the joint solution of Eqs (\ref{eq_particles_N_conserv}, \ref{eq_energy_conserv}) is
\begin{equation}\label{Si_N>s}
    {S}_{I} = p_{s}
\end{equation}
If $N < s$ the family of hyperplanes defined by Eq.(\ref{eq_particles_N_conserv}) does not fully cover the energy hyperplane defined by Eq.(\ref{eq_energy_conserv}) so that the total number of states ${S}_{I}$ corresponding to the joint solution of Eqs (\ref{eq_particles_N_conserv}, \ref{eq_energy_conserv}) obeys
\begin{equation}\label{Si_N<s}
    {S}_{I} < p_s
\end{equation}
In this case no close formula for the number of states ${S}_{I}$ has been obtained. However, it must be stressed that since the particles are distinguishable, it is not the number of \textit{states}, but the number of \textit{configurations} that forms the statistical basis of the present system. In particular, there are
\begin{equation}\label{Cint}
    {C} (n_{0}, n_{1}, n_{2}, \cdots, n_{s}) = \frac{N!}{n_{0}!n_{1}!n_{2}!\cdots n_{s}!}
\end{equation}
configurations corresponding to each state $(n_{0}, n_{1}, n_{2}, \cdots, n_{s})$.

According to Boltzmann's principle: The probability that a state will occur is given by 
\begin{equation}\label{probint}
    p(n_{0}, n_{1}, n_{2}, \cdots, n_{s}) = \frac{1}{{C}_{I}}{C}(n_{0}, n_{1}, n_{2}, \cdots, n_{s})
\end{equation}
where ${{C}_I}$ is the total number of configurations. On this basis, we derive explicitly in Section \ref{sectionI} the probability $p(\kappa)$ that a particle is in the $\kappa$-th energy level i.e contains $\kappa$ quanta. This part of the calculation is developed according to the concept of the \textit{average state} \cite{DarwinFowler} describing the system and it will be shown by considering the special case ($N=4$; $s=4$) that the alternative concept of \textit{most probable state} used in the literature \cite{Boltzmann} is not valid for finite values of $s$ and $N$.

Next, we prove that the above conventional theory gives results that are identical to the ones obtained by considering the distribution of quanta among particles according to the equation
\begin{equation}\label{eq_k_cons_quantat}
\kappa_1 + \kappa_2 + \kappa_3 + \cdots + \kappa_N = s
\end{equation}
where $\kappa_{1} \geq 0,\kappa_{2} \geq 0, \kappa_{3} \geq 0, \cdots, \kappa_{N} \geq 0$ are the respective numbers of quanta attributed to the $N$ particles, defining a state ($\kappa_{1}, \kappa_{2}, \cdots, \kappa_{N}$) of the system. However, it is demonstrated in Section \ref{sectionII} that this identity of results is valid only if the quanta obey a principle of \textit{equal probabilities of states} rather than \textit{configurations}, which leads to the conclusion that the quanta are \textit{indistinguishable} and that the Bose statistics \cite{Ladau} is valid.
As it is well know from the simple problem of distributing $s$ balls into $N$ boxes, the number of states (or solutions of Eq.(\ref{eq_k_cons_quantat})) is given by

\begin{equation}\label{S_II_first}
{S}_{II} = \frac{(N+s-1)!}{s!(N-1)!}
\end{equation}

Note that Eq.(\ref{S_II_first}) will be proved rigorously in Section \ref{sectionII}.

The mathematical method used throughout the article is the Darwin-Fowler technique of generating functions \cite{DarwinFowler}\cite{terHaar} that are appropriate for the study of both approaches.

In the final part of the paper we discuss the passage to the classical limit $\epsilon_{o} \rightarrow 0$; $s \rightarrow \infty$; $s \epsilon_{o} = E$ where the particles have continuous energies. In this case the principle of equal probabilities of quantum states introduced through Eq.(\ref{eq_k_cons_quantat}) transforms to the principle of \textit{equal a priori probabilities} of classical statistical mechanics, justifying the latter principle as a limiting hypothesis based on quantum mechanics.
\vspace{1em}
\subsubsection*{Simple Example}\label{examplesection}

Before developing the general theory, let us present a simple example in order to explain the main idea of the paper. Consider a system of $N=4$ distinguishable particles containing $s=4$ quanta of equal energy.\newline
\textbf{I.} According to the Boltzmann method\cite{Boltzmann}, the statistical analysis of this system is based on Eqs (\ref{eq_particles_N_conserv}, \ref{eq_energy_conserv}) which in the present case have the form: \begin{align}
\label{eq_particles_N=3_conserv} & n_0 + n_1 + n_2 + n_3 + n_4 = 4\\
\label{eq_energly_conserv_s=3} & n_1 + 2 n_2 + 3 n_3 + 4 n_4 = 4
\end{align}
where $n_{0},n_{1}, n_{2}, n_{3}, n_{4}$ represent the number of particles occupying the energy levels $\kappa = 0, 1, 2, 3, 4$ respectively. We observe that there are $S_I = 5$ states of this system (equal to the number of non-negative solutions of Eqs (\ref{eq_particles_N=3_conserv}, \ref{eq_energly_conserv_s=3})) given in Table \ref{table_n}. %partitions of the number $3$) which satisfy both Eqs (\ref{eq_particles_N=3_conserv}, \ref{eq_energly_conserv_s=3}).

\begin{table}[htbp!]
\centering 
\begin{tabular}{| c | c | c | c | c |} 
 \hline
 $n_0$ & $n_1$ & $n_2$ & $n_3$ & $n_4$\\ [0.5ex] 
 \hline\hline
 0 & 4 & 0 & 0 & 0\\ [1ex] 
   \hline
 1 & 2 & 1 & 0 & 0\\ [1ex] 
   \hline
 2 & 0 & 2 & 0 & 0\\ [1ex] 
\hline
 2 & 1 & 0 & 1 & 0\\ [1ex] 
\hline
 3 & 0 & 0 & 0 & 1\\ [1ex] 
\hline
 \end{tabular}
\caption {The 5 states of the system $N=4$; $s=4$.}
\label{table_n}
\end{table}
\vspace{-1em}
\FloatBarrier
%\vspace{-1em}%em}
%\setlength{\belowcaptionskip}{-12pt}
Explicitly, according to Eqs (\ref{partition_ps_formula}, \ref{Si_N>s}) the number of states is equal to the number of partitions $p_4$ of number $4$:
\begin{equation}
    {S}_{I} = p_4 = \frac{2}{\pi}  \int_{0}^{\frac{\pi}{2}} \frac{sin(5x)sin(6x)sin(7x)sin(8x)}{sin(x)sin(2x)sin(3x)sin(4x)} cos(8x) \; dx = 5
\end{equation}
From Eq.(\ref{Cint}) the number of configurations for each state of Table \ref{table_n} reads:
\begin{equation}\label{conf_example}
\begin{split}
{C} (0, 4, 0, 0, 0) = \frac{4!}{0!4!0!0!0!} = 1\\
{C} (1, 2, 1, 0, 0) = \frac{4!}{1!2!1!0!0!} = 12\\
{C} (2, 0, 2, 0, 0) = \frac{4!}{2!0!2!0!0!} = 6\\
{C} (2, 1, 0, 1, 0) = \frac{4!}{2!1!0!1!0!} = 12\\
{C} (3, 0, 0, 0, 1) = \frac{4!}{3!0!0!0!1!} = 4
\end{split}
\end{equation}
and the total number of configurations is: 
\begin{equation}\label{example_3_C_T}
{C}_{I}=1+12+6+12+4=35
\end{equation}
According to the Boltzmann principle, all configurations have equal probability of occurrence.
Therefore from Eq.(\ref{probint}), the corresponding probabilities of the five states of Table \ref{table_n} are given by:
\begin{IEEEeqnarray}{rCl}\label{example_prob}
%a & = & b + c \\
p(0, 4, 0, 0, 0) & = & \frac{1}{35}\;;\;\;p(1, 2, 1, 0, 0)= \frac{12}{35}\;;\;\;p(2, 0, 2, 0, 0)=\frac{6}{35}\nonumber\\
\nonumber\\
p(2, 1, 0, 1, 0) & = & \frac{12}{35}\;;\;\;p(3, 0, 0, 0, 1) = \frac{4}{35} \IEEEyesnumber
\end{IEEEeqnarray}

Next, we define $p(n/\kappa)$ to be the conditional probability that there are $n$ particles in the $\kappa$-th energy level. We calculate $p(n/\kappa)$ for each particular level $\kappa = 0, 1, 2, 3, 4$ using the probabilities of Eqs (\ref{example_prob}) as follows:

Energy level $\kappa=0$:
\begin{equation}\label{example_k_0}
p(0/0)=\frac{1}{35}\;;\;\;p(1/0)=\frac{12}{35}\;; \;\;p(2/0)=\frac{6}{35}+\frac{12}{35}=\frac{18}{35}\;;\;\;p(3/0)=\frac{4}{35}\;;\;\;p(4/0)=0
\end{equation}
Normalisation: $p(0/0) + p(1/0) + p(2/0) + p(3/0) + p(4/0)=1$.\newline
Average number of particles in level $\kappa=0$:
\begin{equation}\label{example_n0}
\langle n_0 \rangle = 0 \cdot p(0/0) + 1 \cdot p(1/0) + 2 \cdot p(2/0) + 3 \cdot p(3/0) + 4 \cdot p(4/0) = \frac{60}{35}
\end{equation}

Energy level $\kappa=1$:
\begin{equation}\label{example_k_1}
p(0/1)=\frac{6}{35}+\frac{4}{35}=\frac{10}{35}\;;\;\;p(1/1)=\frac{12}{35}\;; \;\;p(2/1)=\frac{12}{35}\;;\;\;p(3/1)=0\;;\;\;p(4/1)=\frac{1}{35}
\end{equation}
Normalisation: $p(0/1) + p(1/1) + p(2/1) + p(3/1) + p(4/1)=1$.\newline
Average number of particles in level $\kappa=1$:
\begin{equation}\label{example_n1}
\langle n_1 \rangle = 0 \cdot p(0/1) + 1 \cdot p(1/1) + 2 \cdot p(2/1) + 3 \cdot p(3/1) + 4 \cdot p(4/1)= \frac{40}{35}
\end{equation}

Energy level $\kappa=2$:
\begin{equation}\label{example_k_2}
p(0/2)=\frac{1}{35}+\frac{12}{35}+\frac{4}{35}=\frac{17}{35}\;;\;\;p(1/2)=\frac{12}{35}\;; \;\;p(2/2)=\frac{6}{35}\;;\;\;p(3/2)=0\;;\;\;p(4/2)=0
\end{equation}
Normalisation: $p(0/2) + p(1/2) + p(2/2) + p(3/2) + p(4/2)=1$.\newline
Average number of particles in level $\kappa=2$:
\begin{equation}\label{example_n2}
\langle n_2 \rangle = 0 \cdot p(0/2) + 1 \cdot p(1/2) + 2 \cdot p(2/2) + 3 \cdot p(3/2) + 4 \cdot p(4/2) = \frac{24}{35}
\end{equation}

Energy level $\kappa=3$:
\begin{equation}\label{example_k_3}
p(0/3)=\frac{1}{35}+\frac{12}{35}+\frac{6}{35}+\frac{4}{35} = \frac{23}{35}\;;\;\;p(1/3)=\frac{12}{35}\;; \;\;p(2/3)=0\;;\;\;p(3/3)=0\;;\;\;p(4/3)=0
\end{equation}
Normalisation: $p(0/3) + p(1/3) + p(2/3) + p(3/3) + p(4/3)=1$.\newline
Average number of particles in level $\kappa=3$:
\begin{equation}\label{example_n3}
\langle n_3 \rangle = 0 \cdot p(0/3) + 1 \cdot p(1/3) + 2 \cdot p(2/3) + 3 \cdot p(3/3) + 4 \cdot p(4/3) = \frac{12}{35}
\end{equation}

Energy level $\kappa=4$:
\begin{equation}\label{example_k_4}
p(0/4)=\frac{1}{35}+\frac{12}{35}+\frac{6}{35}+\frac{12}{35} = \frac{31}{35}\;;\;\;p(1/4)=\frac{4}{35}\;; \;\;p(2/4)=0\;;\;\;p(3/4)=0\;;\;\;p(4/4)=0
\end{equation}
Normalisation: $p(0/4) + p(1/4) + p(2/4) + p(3/4) + p(4/4)=1$.\newline
Average number of particles in level $\kappa=4$:
\begin{equation}\label{example_n4}
\langle n_4 \rangle = 0 \cdot p(0/4) + 1 \cdot p(1/4) + 2 \cdot p(2/4) + 3 \cdot p(3/4) + 4 \cdot p(4/4) = \frac{4}{35}
\end{equation}

As expected
\begin{equation}\label{example_sum_n0-n4}
\langle n_0 \rangle + \langle n_1 \rangle + \langle n_2 \rangle + \langle n_3 \rangle + \langle n_4 \rangle = \frac{60 + 40 + 24 + 12 + 4}{35} = 4
\end{equation}

\begin{figure}[!htbp]
\centering
\includegraphics[width=0.7\textwidth]{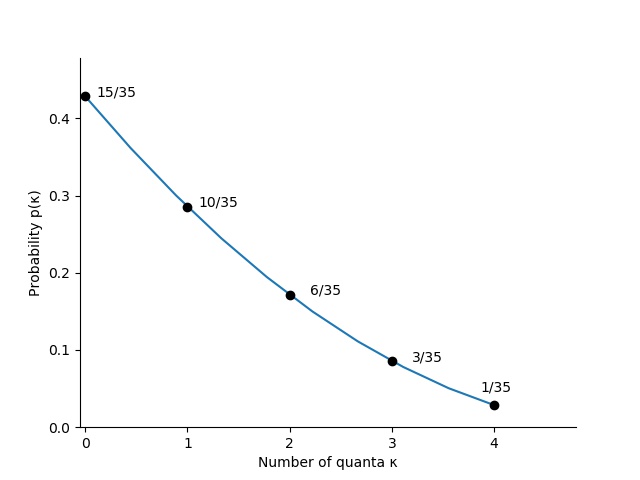}
\caption{Example $N=4$; $s=4$: Graph of the probability that a particle is in the $\kappa$-th energy level.}
\label{pict_1}
\end{figure}

From the above analysis we can define the \textit{average probability} $p(\kappa)$ that a particle is in the $\kappa$-th energy level or equivalently that this particle contains $\kappa$ quanta of energy:

\begin{equation}\label{p_k}
p(\kappa)=\frac{\langle n_{\kappa} \rangle}{N} \;\;\;;\;\;\; \kappa = 0, 1, 2, \cdots, s
\end{equation}
In the present case, according to  the results of the Eqs (\ref{example_n0}-\ref{example_n4}), Eq.(\ref{p_k}) gives for $p(\kappa)$ the values:
\begin{equation}\label{example_probabilities}
p(0)=\frac{15}{35}\;;\;\;p(1)=\frac{10}{35}\;; \;\;p(2)=\frac{6}{35}\;;\;\;p(3)=\frac{3}{35}\;; \;\;p(4)=\frac{1}{35}
\end{equation}
represented in Figure \ref{pict_1}.

Normalisation: \begin{equation}\label{example_norm}p(0) + p(1) + p(2) + p(3) + p(4)=1\end{equation}
The average number of quanta existing in a particle is:

\begin{equation}\label{example_av_quanta_in_particle}
\langle \kappa \rangle = 1 \cdot \frac{10}{35} + 2 \cdot \frac{6}{35} + 3 \cdot \frac{3}{35} + 4 \cdot \frac{1}{35} = 1
\end{equation}
which is consistent with $\langle \kappa \rangle = s/N$.
\vspace{-1em}
\FloatBarrier

We notice here that many authors, including Boltzmann \cite{Boltzmann} define $p(\kappa)$ using the concept of \textit{most probable state} instead of Eq.(\ref{p_k}). According to this concept, one would derive the state (${n^*_0}, {n^*_1}, {n^*_2}, \cdots, {n^*_s}$) that maximizes the number of configurations ${C} (n_0, n_1, \cdots, n_s)$ given by Eq.(\ref{Cint}) and he would define:
\begin{equation} \label{not_unique_p_k}
    p(\kappa) = \frac{{n^*_{\kappa}}}{N}\;\;\;;\;\;\; \kappa=0, 1, 2, \cdots, s
\end{equation}

We observe from the present example, however, that this is not a valid argument in general because the number of configurations ${C}(n_0,n_1,n_2,n_3,n_4)$ given by Eqs (\ref{conf_example}) has two different maxima ${C}=12$ at (${n^*_0}=1, {n^*_1}=2, {n^*_2}=1, {n^*_3}=0, {n^*_4}=0$) and (${n^*_0}=2, {n^*_1}=1, {n^*_2}=0, {n^*_3}=1, {n^*_4}=0$). Therefore $p(\kappa)$ defined by Eq.(\ref{not_unique_p_k}) is not unique and the concept of most probable configuration cannot be used for finite $N$ and $s$. A rather general criticism of this approach can also be found in the paper of Darwin and Fowler \cite{DarwinFowler}.

\vspace{5mm} %5mm vertical space

\textbf{II.} Let us next distribute the energy quanta over the particles according to Eq.(\ref{eq_k_cons_quantat}) which in the present case has the form:
\begin{equation}\label{eq_k_4_cons_quanta}
\kappa_1 + \kappa_2 + \kappa_3 + \kappa_4 = 4
\end{equation}
where $\kappa_{1},\kappa_{2}, \kappa_{3}, \kappa_{4}$ are respectively the number of quanta contained in the particles $n = 1, 2, 3, 4$.
According to Eq.(\ref{S_II_first}), the total number of states (non-negative integer solutions of Eq.(\ref{eq_k_4_cons_quanta})) is 
\begin{equation}\label{ST_4}
S_{II} = \frac{(4+4-1)!}{4! (4-1)!} = 35
\end{equation}

States $S_{II}$ (Eq.(\ref{ST_4})) are represented by the points of the non-negative grid in 4-D space [$\kappa_1, \kappa_2, \kappa_3, \kappa_4$] that also belong to the hyperplane defined by Eq.(\ref{eq_k_4_cons_quanta}).

From Eqs (\ref{example_3_C_T}, \ref{ST_4}) we notice the very important equality ${C}_{I} = S_{II}$ which will be proved to be valid in general later in the paper.

Assuming that the states $S_{II}$ in Eq.(\ref{ST_4}) have equal probability of occurrence means that the quanta are \textit{indistinguishable} and that the probability $p(\kappa)$ that any one of the four particles (say particle $1$) has $\kappa$ quanta, can be calculated by re-writing Eq.(\ref{eq_k_4_cons_quanta}) in the form $\kappa_2 + \kappa_3 + \kappa_4 = 4 - \kappa_1$ where $\kappa_1 = 0, 1, 2, 3, 4$. %$\kappa_i = 0, 1, 2, 3, 4$.
\begin{equation}\label{example_k1}
  \begin{split}
\kappa_{1} = 0 :\;\; \kappa_{2}+\kappa_{3}+\kappa_{4}=4;\;\; S_{0}=\frac{(4+3-1)!}{4! (3-1)!}= \text{15 states};\;\;p(0)=\frac{S_{0}}{S_{II}}=\frac{15}{35}
\\\
\kappa_{1} = 1 :\;\; \kappa_{2}+\kappa_{3}+\kappa_{4}=3;\;\; S_{1}=\frac{(3+3-1)!}{3! (3-1)!}= \text{10 states};\;\;p(1)=\frac{S_{1}}{S_{II}}=\frac{10}{35}
\\\
\kappa_{1} = 2 :\;\; \kappa_{2}+\kappa_{3}+\kappa_{4}=2;\;\; S_{2}=\frac{(2+3-1)!}{2! (3-1)!}= \text{6 states};\;\;p(2)=\frac{S_{2}}{S_{II}}=\frac{6}{35}
\\\
\kappa_{1} = 3 :\;\; \kappa_{2}+\kappa_{3}+\kappa_{4}=1;\;\; S_{3}=\frac{(1+3-1)!}{1! (3-1)!}= \text{3 states};\;\;p(3)=\frac{S_{3}}{S_{II}}=\frac{3}{35}
\\\
\kappa_{1} = 4 :\;\; \kappa_{2}+\kappa_{3}+\kappa_{4}=0\;;\;\; S_{4}=\frac{(0+3-1)!}{0! (3-1)!}= \text{1 state}\;;\;\;p(4)=\frac{S_{4}}{S_{II}}=\frac{1}{35}
\end{split}
\end{equation}
We observe that results of Eqs (\ref{example_k1}) coincide with Eqs (\ref{example_probabilities}). This remarkable conclusion i.e. that $p(\kappa)$ can be obtained both by distributing \textit{distinguishable} particles over energy levels or by distributing \textit{indistinguishable} quanta over particles, will be proved to be valid in general in the next sections of the paper. An exact formula for $p(\kappa)$ will be derived for all finite values of $N$ and $s$ and the limit $N\rightarrow \infty $; $s\rightarrow \infty$;  $s/N= \langle \kappa \rangle >> 1$ will be studied.

\section{Distribution of particles in energy levels}\label{sectionI}

The total number of configurations of $N$ particles occupying the energy levels $\kappa = 0, 1, 2, \cdots, s$ is given according to Eqs (\ref{eq_particles_N_conserv}, \ref{eq_energy_conserv}) by
\begin{equation}\label{C_T}
{C}_{I} = \sum_{n_{0}=0}^{N}\sum_{n_{1}=0}^{N} \cdots \sum_{n_{s}=0}^{N} \frac{N!}{n_{0}! n_{1}! \cdots n_{s}!} \; \delta(n_0 + n_1 + n_2 + \cdots + n_s - N) \; \delta( n_1 + 2 n_2 + \cdots + s n_s - s)
\end{equation}

where 
\begin{equation}\label{delta_kronecker}
\delta( m - n ) = \begin{cases}
1 & \text{; $m = n$}\\
0 & \text{; $m \neq n$}\\
\end{cases}
\end{equation}

The generating function in this case is

\begin{equation}\label{gen_function_f(x,y)}
f(x, y) =   \sum_{n_{0}=0}^{\infty}\sum_{n_{1}=0}^{\infty} \cdots \sum_{n_{s}=0}^{\infty}  \frac{N!}{n_{0}! n_{1}! \cdots n_{s}!} \; x^{n_{0}+n_{1}+\cdots+n_{s}} \; y^{n_{1}+2 n_{2}+\cdots+s n_{s}}
\end{equation}

where $0 \leq x < 1;\;\; 0 \leq y < 1$. 
Taking separate factors in Eq.(\ref{gen_function_f(x,y)}) we have
\begin{equation}\label{sep_factors}
\sum_{n_{0}=0}^{\infty} \frac{x^{n_{0}}}{n_{0}!} = e^x ; \;\; \sum_{n_{1}=0}^{\infty} \frac{(x y)^{n_1}}{n_{1}!} = e^{xy} ; \;\; \sum_{n_{2}=0}^{\infty} \frac{(x y^2)^{n_2}}{n_{2}!} = e^{x {y}^{2}} ; \;\;\cdots \;\; \sum_{n_{s}=0}^{\infty}  \frac{(x y^s)^{n_s}}{n_{s}!}= e^{x{y}^{s}}
\end{equation}

and the generating function becomes
\begin{equation*}%\label{gen_fun_calc}
f(x, y) = N! \; e^{x} \cdot e^{xy} \cdot e^{x {y}^{2}} \cdots e^{x {y}^{s}} = N! \exp \big\{ x (1 + y + y^{2} + \cdots + y^s) \big\}    
\end{equation*}
so that
\begin{equation}\label{gen_fun_exp}
f(x, y) = N! \exp \Big\{x \Big(\frac{1- y^{s+1}}{1-y}\Big) \Big\} 
%\exp \big\{ x (1 + y + y^{2} + \cdots + y^s) \big\}    
\end{equation}

Let us expand $f(x, y)$ in powers of $x$:
\begin{equation}\label{gen_fun_expand_powers}
f(x, y) = N! \Big\{ 1 + x \frac{1- y^{s+1}}{1-y} + \frac{x^2}{2!} \Big(\frac{1- y^{s+1}}{1-y}\Big)^2 + \cdots \Big\} 
\end{equation}

the coefficient of $x^N$ in this expansion is
\begin{equation}\label{coeff_g(y)}
g(y) = \Big( \frac{1- y^{s+1}}{1-y}  \Big)^N
\end{equation}

We argue that if $g(y)$ is expressed in the form of a power series in $y$, then the coefficient of $y^s$ coincides with ${C}_I$ defined by Eq.(\ref{C_T}):

\begin{equation}\label{C_T_deriv}
{C}_{I} =\frac{1}{s!}\Bigg[\frac{\partial ^s}{\partial y^s} \Big(\frac{1-y^{s+1}}{1-y} \Big)^{N} \Bigg]_{y=0}     
\end{equation}

We observe that $(1 - y^{s+1})^{N} \approx 1 - N y^{s+1} + \mathcal{O} [y^{2(s+1)}]$ and therefore
\begin{IEEEeqnarray}{rCl}\label{calculations_of_C_T}
%a & = & b + c \\
\Bigg[ \frac{\partial ^s}{\partial y^s} \Big( \frac{1 - y^{s+1}}{1-y} \Big)^{N} \Bigg]_{y = 0} & = & \Bigg[ \frac{\partial ^s}{\partial y^s} \frac{1}{(1-y)^N} \Bigg]_{y = 0}\nonumber\\
% & = & d + e \IEEEyesnumber
& = & \Bigg[ \frac{N(N+1)(N+2)\cdots (N+s-1)}{(1-y)^{N+s}} \Bigg]_{y=0} = \frac{(N+s-1)!}{(N-1)!}\IEEEyesnumber
\end{IEEEeqnarray}
and Eq.(\ref{C_T_deriv}) gives the total number of configurations of all states:

\begin{equation}\label{C_T_fraction}
{C}_{I} = \frac{(N+s-1)!}{s!(N-1)!}
\end{equation}

According to the Boltzmann principle, all configurations have equal probability of occurrence, therefore the probability that state $(n_{0}, n_{1}, n_{2}, \cdots, n_{s})$ will occur is
\begin{equation}\label{prob_of_state}
p(n_{0}, n_{1}, n_{2}, \cdots, n_{s}) = \frac{1}{{C}_{I}} \frac{N!}{n_{0}! n_{1}! \cdots n_{s}!} \; \delta (n_{0} + n_{1}+ \cdots + n_{s} -N) \; \delta(n_{1} + 2n_{2} + \cdots + s n_{s} -s)   
\end{equation}

The average number of particles located in level $\kappa$ is
\begin{IEEEeqnarray}{rCl}\label{meso_n_k}
%a & = & b + c \\
\langle n_{\kappa} \rangle & = & \sum_{n_{0}=0}^{N}\sum_{n_{1}=0}^{N} \cdots \sum_{n_{s}=0}^{N} n_{\kappa} \; p(n_{0}, n_{1}, n_{2}, \cdots, n_{s})\nonumber\\
% & = & d + e \IEEEyesnumber
& = & \frac{N!}{{C}_I} \sum_{n_{0}=0}^{N}\sum_{n_{1}=0}^{N} \cdots \sum_{n_{s}=0}^{N} \frac{n_{\kappa}}{n_{0}! n_{1}! \cdots n_{s}!} \; \delta (n_{0} + n_{1}+ \cdots + n_{s} -N) \; \delta(n_{1} + 2n_{2} + \cdots + s n_{s} -s)\IEEEyesnumber
\end{IEEEeqnarray}
Generating function:
\begin{equation}\label{gen_func_4}
f_{\kappa}(x,y)=\frac{N!}{{C}_{I}}\sum_{n_{0}=0}^{\infty}\sum_{n_{1}=0}^{\infty}\cdots \sum_{n_{s}=0}^{\infty}\frac{n_{\kappa}}{n_{0}!n_{1}! \cdots n_{s}!} \; x^{n_{0}+n_{1}+\cdots+n_{s}} \; y^{n_{1}+2n_{2}+\cdots+sn_{s}}
\end{equation}

where $0 \leq x < 1;\;\; 0 \leq y < 1$. 
Taking separate factors we have
\begin{IEEEeqnarray}{rCl}\label{eq_a_b}
 \sum_{n_{i}=0}^{\infty} & \frac{(x {y}^{i})^{n_{i}}}{n_{i}!} & = e^{x {y}^{i}} \;\;\;;\;\;\; i=0, 1, 2,\cdots, \kappa -1, \kappa +1, \cdots, s  \IEEEyesnumber\IEEEyessubnumber\\
 \sum_{n_{\kappa}=0}^{\infty} & n_{\kappa} \frac{(x{y}^{\kappa})^{n_{\kappa}}}{n_{\kappa}!} & = x{y}^{\kappa} \sum_{n_{\kappa}=1}^{\infty} \frac{(x{y}^{\kappa})^{n_{\kappa}-1}}{(n_{\kappa}-1)!} = x{y}^{\kappa} e^{x{y}^{\kappa}}\IEEEyessubnumber
\end{IEEEeqnarray}
so that
\begin{IEEEeqnarray}{rCl}\label{gen_func_5}
f_{\kappa}(x, y) & = & \frac{N!}{{C}_{I}} x{y}^{\kappa} \exp \{ x (1+y+y^{2}+\cdots+ y^{s}) \} \nonumber\\
& = & \frac{N!}{{C}_{I}} x{y}^{\kappa} 
\exp \Big\{ x \Big( \frac{1-{y}^{s+1}}{1-y}\Big) \Big\}\IEEEyesnumber
\end{IEEEeqnarray}

Let us expand $f_{\kappa}(x, y)$ in powers of $x$:
\begin{equation}\label{gen_func_6}
f_{\kappa}(x, y) = \frac{N!}{{C}_{I}} x{y}^{\kappa} \Big\{ 1+ x \frac{1- {y}^{s+1}}{1-y} + \frac{x^2}{2!} \Big( \frac{1- {y}^{s+1}}{1-y} \Big)^{2} + \cdots  \Big\}
\end{equation}

the coefficient of $x^N$ in this expansion is
\begin{equation}\label{gen_func_7}
g_{\kappa}(y) = \frac{N}{{C}_{I}} {y}^{\kappa} \Big( \frac{1-{y}^{s+1}}{1-y} \Big)^{N-1}
\end{equation}

We argue that if $g_{\kappa}(y)$ is expressed in the form of a power series in $y$, then the coefficient of $y^s$ coincides with $\langle n_{\kappa} \rangle$ defined by Eq.(\ref{meso_n_k}):
\begin{equation}\label{meso_n_k_deriv}
\langle n_{\kappa} \rangle  = \frac{N}{{C}_{I} s!} \Bigg[ \frac{\partial ^s}{\partial y^s} \Big\{ y^{\kappa} \Big( \frac{1 - y^{s+1}}{1-y} \Big)^{N-1} \Big\} \Bigg]_{y = 0} 
\end{equation}

We observe that $(1- {y}^{s+1})^{N-1} \approx 1- (N-1) y^{s+1} + \mathcal{O} [y^{2(s+1)}]$ and therefore

\begin{equation}\label{meso_n_k_deriv_2}
\langle n_{\kappa} \rangle  = \frac{N}{{C}_{I} s!} \Bigg[ \frac{\partial ^s}{\partial y^s} \Bigg(
\frac{y^{\kappa}}{{(1-y)}^{N-1}}\Bigg) \Bigg]_{y = 0}
\end{equation}

Let us next use the binomial expansion
\begin{equation}\label{bin_exp}
\frac{\partial ^s}{\partial y^s} \Bigg\{
\frac{y^{\kappa}}{{(1-y)}^{N-1}}\Bigg\} = \sum_{l=1}^{s} \begin{pmatrix} s \\ l \end{pmatrix} \frac{\partial ^l}{\partial y^l} \Bigg\{ \frac{1}{{(1-y)}^{N-1}} \Bigg\} \frac{\partial ^{s-l}}{\partial y^{s-l}} \big(y^{\kappa}\big)
\end{equation}

where
\begin{IEEEeqnarray}{rCl}\label{deriv_calc}
\frac{\partial ^l}{\partial y^l} \Bigg\{ \frac{1}{{(1-y)}^{N-1}} \Bigg\} & = & \frac{(N-1)N(N+1)\cdots (N+l-2)}{{(1-y)}^{N+l-1}} \IEEEyesnumber \IEEEyessubnumber\\
\frac{\partial ^{s-l}}{\partial y^{s-l}} \big(y^{\kappa}\big) & = & \begin{cases}
\frac{\kappa !}{(\kappa - s+l)!} y^{\kappa - s+l} & \text{if $l \geq s - \kappa$}\\
0 & \text{if $l < s - \kappa$}\\
\end{cases}
\IEEEyessubnumber
\end{IEEEeqnarray}

Substituting the latter results into the binomial expansion of Eq.(\ref{bin_exp}) we get

\begin{equation}\label{}
\frac{\partial ^s}{\partial y^s} \Bigg\{
\frac{y^{\kappa}}{{(1-y)}^{N-1}}\Bigg\} = \sum_{l=s-\kappa}^{s} \begin{pmatrix} s \\ l \end{pmatrix} \frac{(N+l-2)!}{(N-2)!} \cdot \frac{1}{{(1-y)}^{N+l-1}} \cdot \frac{\kappa !}{(\kappa -s+l)!}\; {y}^{\kappa -s+l}
\end{equation}
so that at $y=0$ only the term $y^0$ is not zero corresponding to $l=s-\kappa$. Therefore $\langle n_{\kappa} \rangle$ can be obtained from Eq.(\ref{meso_n_k_deriv_2}) as
\begin{equation}\label{meso_n_k_matrix}
\langle n_{\kappa} \rangle = \frac{N}{{C}_{I}} \begin{pmatrix} N+s-\kappa -2 \\ N-2 \end{pmatrix} \;\;\; ; \;\;\; \kappa = 0, 1, 2, \cdots, s
\end{equation}

where ${C}_{I}$ is given by Eq.(\ref{C_T_fraction}). Note that Eq.(\ref{meso_n_k_matrix}) has also been derived by Darwin and Fowler \cite{DarwinFowler}. \newline The probability that a particle is in the $\kappa$-th energy level i.e. contains $\kappa$ quanta is
\begin{equation}\label{probability_p(k)}
p(\kappa) = \frac{\langle n_{\kappa} \rangle}{N} = \frac{\begin{pmatrix} N+s-\kappa -2 \\ N-2 \end{pmatrix}}{\begin{pmatrix} N+s -1 \\ N-1 \end{pmatrix}} \;\;\; ; \;\;\; \kappa = 0, 1, 2, \cdots, s
\end{equation}

In the special case $N=4$; $s=4$ considered in the introduction we obtain
\begin{equation}\label{probability_p(k=3)}
p(\kappa) = \frac{\begin{pmatrix} 6-\kappa \\ 2 \end{pmatrix}}{\begin{pmatrix} 7 \\ 3 \end{pmatrix}} = \frac{(5-\kappa)(6-\kappa)}{70} \;\;\; ; \;\;\; \kappa = 0, 1, 2, 3, 4
\end{equation}

which reproduces exactly the results of Eqs (\ref{example_probabilities}).

\section{Distribution of energy quanta among particles}\label{sectionII}

The total number of states of $s$ quanta distributed among $N$ particles is given according to Eq.(\ref{eq_k_cons_quantat}) by
\begin{equation}\label{eq_S_T_delta_func}
{S}_{II} = \sum_{{\kappa}_{1}=0}^{s}\sum_{{\kappa}_{2}=0}^{s} \cdots \sum_{{\kappa}_{N}=0}^{s} \delta ({\kappa}_{1} + {\kappa}_{2}+ \cdots + {\kappa}_{N} -s)
\end{equation}

The generating function in this case is
\begin{equation}\label{gen_function_section_3}
F(x) = \sum_{{\kappa}_{1}=0}^{\infty}\sum_{{\kappa}_{2}=0}^{\infty} \cdots \sum_{{\kappa}_{N}=0}^{\infty} \; x^{{\kappa}_{1} + {\kappa}_{2}+ \cdots + {\kappa}_{N}}
\end{equation}

where $0 \leq x < 1$. Taking separate factors we have
\begin{equation}\label{sep_factors_section_3}
\sum_{{\kappa}_{1}=0}^{\infty} x^{{\kappa}_{1}} = \frac{1}{1-x} ; \;\; \sum_{{\kappa}_{2}=0}^{\infty} x^{{\kappa}_{2}} = \frac{1}{1-x} ; \;\;\cdots \;\; \sum_{{\kappa}_{N}=0}^{\infty} x^{{\kappa}_{N}} = \frac{1}{1-x}
\end{equation}

so that
\begin{equation}\label{gen_function_section_3_fraction}
F(x) = \frac{1}{(1-x)^N}
\end{equation}

Let us expand $F(x)$ in powers of $x$:
\begin{equation}\label{gen_function_section_3_expansion}
F(x) = 1 + N x + \frac{N(N+1)}{2!}\;x^2 + \cdots
\end{equation}

We argue that the coefficient of $x^s$ in series (\ref{gen_function_section_3_expansion}) coincides with $S_{II}$ defined by Eq.(\ref{eq_S_T_delta_func}):
\begin{equation}\label{S_T_deriv}
{S}_{II} =\frac{1}{s!}\Bigg[\frac{\partial ^s}{\partial x^s} \Bigg(\frac{1}{(1-x)^N} \Bigg) \Bigg]_{x=0}     
\end{equation}

We have
\begin{equation}\label{calculations_of_S_T}
\Bigg[ \frac{\partial ^s}{\partial x^s} \Big\{ \frac{1}{(1-x)^N} \Big\} \Bigg]_{x = 0} = \Bigg[ \frac{N(N+1)(N+2)\cdots (N+s-1)}{(1-x)^{N}} \Bigg]_{x = 0} = \frac{(N+s-1)!}{(N-1)!}
\end{equation}

and Eq.(\ref{S_T_deriv}) gives the total number of quantum states in accordance with Eq.(\ref{S_II_first}):

\begin{equation}\label{S_T_fraction}
{S}_{II} = \frac{(N+s-1)!}{s!(N-1)!}
\end{equation}

From Eq.(\ref{C_T_fraction}) and Eq.(\ref{S_T_fraction}) we observe that the important equality between the number of configurations in case I (Section \ref{sectionI}) and the number of states in case II (Section \ref{sectionII})
\begin{equation}\label{equality_C_T_S_T}
{C}_{I} = {S}_{II}
\end{equation}
is valid in general for systems of particles and quanta.

The probability $p(\kappa)$ that a particle contains $\kappa$ quanta can be next obtained by introducing the principle that \textit{all quantum states $S_{II}$ which are the non-negative integer solutions of Eq.(\ref{eq_k_cons_quantat}), have equal probability of occurrence}. This means that the quanta are \textit{indistinguishable} and the Bose statistics \cite{Ladau} is valid.

For ${\kappa}_i = \kappa \sim$ fixed, Eq.(\ref{eq_k_cons_quantat}) can be written as
\begin{equation}\label{eq_of_k}
{\kappa}_{1} + {\kappa}_{2} + \cdots + {\kappa}_{i-1} + {\kappa}_{i+1} + \cdots + {\kappa}_{N} = s- \kappa
\end{equation}

and the previous analysis for the derivation of $S_{II}$ can be repeated for Eq.(\ref{eq_of_k}) where $s$ is replaced by $s - \kappa$ and $N$ by $N-1$ so that the number of states where $\kappa_i = \kappa$, is

\begin{equation}\label{S_k_fraction}
{S}_{\kappa} = \frac{(N+s-\kappa -2)!}{(s-\kappa)!(N-2)!}
\end{equation}

and the probability $p(\kappa)$ reads

\begin{equation}\label{probability_p(k)_eq_k}
p(\kappa) = \frac{S_{\kappa}}{S_{II}} = \frac{\begin{pmatrix} N+s-\kappa -2 \\ N-2 \end{pmatrix}}{\begin{pmatrix} N+s -1 \\ N-1 \end{pmatrix}} \;\;\; ; \;\;\; \kappa = 0, 1, 2, \cdots, s
\end{equation}

which is exactly the same as the result of Eq.(\ref{probability_p(k)}). 

The equality of the above results shows the equivalence of the two principles:
\textbf{I.} Equal probability of \textit{configurations} for \textit{distinguishable particles} and \textbf{II.} Equal probability of \textit{states} for \textit{indistinguishable quanta} of energy. These two approaches do not contradict each other but rather present two equally valid perspectives for the statistical analysis of the particle-quantum system.

\section{Properties of the probability distribution}\label{propertiessection}

Distribution $p(\kappa)$ is given by Eqs (\ref{probability_p(k)}, \ref{probability_p(k)_eq_k}) and has the following properties for finite systems.

Normalisation:
\begin{IEEEeqnarray}{rCl}\label{normalisation_of_p(k)}
%a & = & b + c \\
\sum_{{\kappa}=0}^{s} p(\kappa) & = & \frac{1}{\begin{pmatrix} N+s -1 \\ N-1 \end{pmatrix}} \sum_{{\kappa}=0}^{s} \begin{pmatrix} N+s-\kappa -2 \\ N-2 \end{pmatrix} \nonumber\\
% & = & d + e \IEEEyesnumber
& = & \frac{1}{\begin{pmatrix} N+s -1 \\ N-1 \end{pmatrix}} \sum_{l=0}^{s} \begin{pmatrix} N-2+l \\ N-2 \end{pmatrix} = 1 \IEEEyesnumber
\end{IEEEeqnarray}

Average number of quanta existing within a particle:
\begin{IEEEeqnarray}{rCl}\label{average_k}
%a & = & b + c \\
\langle \kappa \rangle = \sum_{{\kappa}=0}^{s} \kappa\;p(\kappa) & = & \frac{1}{\begin{pmatrix} N+s -1 \\ N-1 \end{pmatrix}} \sum_{{\kappa}=0}^{s} \kappa \begin{pmatrix} N+s-\kappa -2 \\ N-2 \end{pmatrix} \nonumber\\
& = & s -(N-1)\frac{s}{N} = \frac{s}{N} \IEEEyesnumber
\end{IEEEeqnarray}
%where $l=s-\kappa$ and $p=l-1$.
Calculation of $\langle \kappa ^2 \rangle$:
\begin{IEEEeqnarray}{rCl}\label{mean_k^2}
& \langle \kappa ^2 \rangle = \sum_{{\kappa}=0}^{s} \kappa^2 \;p(\kappa) = \frac{1}{\begin{pmatrix} N+s -1 \\ N-1 \end{pmatrix}} \sum_{{\kappa}=0}^{s} \kappa^2 \begin{pmatrix} N+s-\kappa -2 \\ N-2 \end{pmatrix} \nonumber\\
& = \frac{1}{\begin{pmatrix} N+s -1 \\ N-1 \end{pmatrix}} \sum_{l=0}^{s} (s-l)^2 \begin{pmatrix} N-2+l \\ N-2 \end{pmatrix} = \frac{N-1}{N+1} \langle \kappa \rangle + \frac{2N}{N+1} \langle \kappa \rangle ^2 \IEEEyesnumber
\end{IEEEeqnarray}
Standard deviation:
\begin{equation} \label{std}
    \sigma ^2 = \langle \kappa ^2 \rangle - \langle \kappa \rangle ^2 = \frac{N-1}{N+1} [\langle \kappa \rangle + \langle \kappa \rangle ^2]
\end{equation}

Let us next rewrite the distribution $p(\kappa)$ given by Eqs (\ref{probability_p(k)}, \ref{probability_p(k)_eq_k}) as follows:%for the limit $s \rightarrow \infty $; $N\rightarrow \infty $.
\begin{IEEEeqnarray}{rCl}\label{p(k)_calc_5}
%a & = & b + c \\
% & = & d + e \IEEEyesnumber
p(\kappa) & = & (N-1)\dfrac{\dfrac{s!}{(s-\kappa)!}}{\dfrac{(N+s-1)!}{(N+s-\kappa-2)!}} \nonumber \\
& = & \dfrac{N-1}{N+s-\kappa-1}\cdot \dfrac{(s-\kappa+1)(s-\kappa+2)\cdots(s-1)s}{(N+s-\kappa)(N+s-\kappa+1)\cdots(N+s-1)} \nonumber \\
& = & \dfrac{1 - \dfrac{1}{N}}{1+\dfrac{s}{N} - \dfrac{\kappa-1}{N}} \cdot \dfrac{s^{\kappa} (1-\dfrac{\kappa-1}{s})(1-\dfrac{\kappa-2}{s})\cdots(1-\dfrac{1}{s})}{(N+s)^{\kappa} (1-\dfrac{\kappa}{N+s})(1-\dfrac{\kappa-1}{N+s})\cdots(1-\dfrac{1}{N+s})}\IEEEyesnumber
\end{IEEEeqnarray}

At the limit $s \rightarrow \infty $; $N\rightarrow \infty $,  $s/N= \langle \kappa \rangle >>1$ we get:
\begin{equation}\label{p(k)limit}
p(\kappa) = \frac{1}{\langle \kappa \rangle} \lim {\Big( 1+\frac{N}{s} \Big)^{-\kappa}} = \frac{1}{\langle \kappa \rangle} \lim \Big[ \Big(1+\frac{N}{s} \Big)^{s}\Big]^{-\mathlarger{\frac{\kappa}{s}}} = \frac{1}{\langle \kappa \rangle} \big(e^{N} \big)^{-\mathlarger{\frac{\kappa}{s}}} = \frac{1}{\langle \kappa \rangle} e^{-\mathlarger{\frac{\kappa}{\langle \kappa \rangle}}}
\end{equation}

From Eq.(\ref{std}) the standard deviation of $p(\kappa)$ at the present limit is $\sigma = \langle \kappa \rangle$.

The classical limit of $p(\kappa)$ where the energies of the particles are continuous can be obtained from Eq.(\ref{p(k)limit}) at the limit $\epsilon_{o} \rightarrow 0$; $s \rightarrow \infty$; $s \epsilon_{o} = E \sim$ fixed. In this case the density of the energy levels increases so that $p(\kappa)d\kappa$ gives with great accuracy the probability that a particle is within an energy level that lies in the interval $[\kappa, \kappa+d\kappa]$. Expressing next the energy $\epsilon$ of a particle as $\epsilon = \kappa \epsilon_{o}$, we can write a probability conservation equation in the form
\begin{equation}\label{p(e)=p(k)}
    P(\epsilon)d\epsilon = p(\kappa)d\kappa
\end{equation}

and we derive the Boltzmann law from Eq.(\ref{p(k)limit}) using $\langle \epsilon \rangle = \epsilon_{o} \langle \kappa \rangle$:
\begin{equation}\label{p(e)}
    P(\epsilon) = \frac{1}{\langle \epsilon \rangle} e^{-\mathlarger{\frac{\epsilon}{\langle \epsilon \rangle}}}
\end{equation}

Note that since Eq.(\ref{p(e)}) is derived from Eq.(\ref{p(k)limit}), it is valid only at the limit $N \rightarrow \infty$; $E \rightarrow \infty$; $E/N = \langle \epsilon \rangle \sim$ fixed.

\section{The classical limit of continuous energies}

Multiplying both sides of Eq.(\ref{eq_k_cons_quantat}) by $\epsilon_o$, and taking the limit $\epsilon_{o} \rightarrow 0$; $s \rightarrow \infty$; $s \epsilon_{o} = E \sim$ fixed, the energies ($\epsilon_{1}, \epsilon_{2}, \cdots, \epsilon_{N}$) of the particles become continuous variables satisfying the equation of the energy hyperplane:
\begin{equation}\label{Heq}
    \mathcal{H}(\epsilon_{1}, \epsilon_{2}, \cdots, \epsilon_{n}) \equiv \epsilon_{1}+ \epsilon_{2} + \cdots + \epsilon_{N} = E 
\end{equation}
where $0 \leq \epsilon_{1} \leq E; \; 0 \leq \epsilon_{2} \leq E; \cdots; \; 0 \leq \epsilon_{N} \leq E$. \par
According to the classical \textit{principle of equal a priori probabilities} for the present closed system: %All points of the positive part of the energy hyperplane are equiprobable. 
The probability that the system is within a region of the energy hyperplane is proportional to the area of that region. It is clear however that in the context of the theory developed already in the present paper, the above principle is not just a hypothesis but it is validated as the classical limit of the principle of equal probabilities of quantum states introduced in Section \ref{sectionII}. It is therefore expected that the Boltzmann law $P(\epsilon)$ derived in Eq.(\ref{p(e)}) as the classical limit of $p(\kappa)$ given by Eq.(\ref{p(k)limit}), may be also calculated directly by projecting the hyperplane of Eq.(\ref{Heq}) over one of the axes where the energy of a single particle is measured.
The surface area $\sigma(\epsilon)$ of the hyperplane of Eq.(\ref{Heq}) can be calculated by considering the differential volume created between the two hyperplanes $\mathcal{H}=E$ and $\mathcal{H}=E+dE$:
\begin{equation}\label{sigma(E)dh}
    \sigma(E)dh = dE  \int_{0}^{\infty}\!\!
\int_{0}^{\infty} \cdots \int_{0}^{\infty} \delta(\epsilon_1 + \epsilon_2 + \cdots + \epsilon_N - E) \; d\epsilon_1 \; d\epsilon_2 \cdots d\epsilon_N
\end{equation}
where the characteristic function here is 
\begin{equation}
    \mathcal{D} = dE \; \delta(\epsilon_1 + \epsilon_2 + \cdots + \epsilon_N - E) =
    \begin{cases}
1 & \text{; $E \leq \epsilon_1 + \epsilon_2 + \cdots + \epsilon_N \leq E + dE $}\\
0 & \text{; otherwise}\\
\end{cases}
\end{equation}

Also we have
\begin{IEEEeqnarray}{rCl}\label{grad_H}
%a & = & b + c \\
& \vec{\nabla}{\mathcal{H}}=  \frac{\partial \mathcal{H}}{\partial \epsilon_1}\hat{i}_1 + & \frac{\partial \mathcal{H}}{\partial \epsilon_2}\hat{i}_2 + \cdots + \frac{\partial \mathcal{H}}{\partial \epsilon_N}\hat{i}_N = \hat{i}_1 + \hat{i}_2 + \cdots + \hat{i}_N \nonumber\\
% & = & d + e \IEEEyesnumber
& \| \vec{\nabla}{\mathcal{H}} \| = \sqrt{N} &\;\;;\;\; dh = \frac{dE}{\sqrt{N}}\IEEEyesnumber
\end{IEEEeqnarray}

Using the Fourier representation of the $\delta$-function we get
\begin{equation}
    \sigma(E) = \sqrt{N} \int_{0}^{+\infty}\!\!
\int_{0}^{+\infty} \cdots \int_{0}^{+\infty} \Big\{ \frac{1}{2\pi}  \int_{-\infty}^{+\infty} e^{i \rho (\epsilon_1 + \epsilon_2 + \cdots + \epsilon_N - E)} \; d\rho \Big\} d\epsilon_1 \; d\epsilon_2 \cdots d\epsilon_N 
\end{equation}

Introducing the parameter $\lambda>0$, the order of the integration can be interchanged and the integrand factorizes as 
\begin{equation}\label{s(E)factor}
    \sigma(E) = \frac{\sqrt{N}}{2 \pi} \lim_{\lambda \rightarrow 0} \int_{-\infty}^{+\infty} e^{-i\rho E} \Big\{ \int_{0}^{+\infty}  e^{-(\lambda-i\rho)\epsilon_1} \; d\epsilon_1 \Big\}^N d\rho
\end{equation}

Separating the integral into real and imaginary parts (see also ref \cite{Integrals_book} p.342):
\begin{equation}\label{integrI}
    I = \int_{0}^{+\infty} e^{-(\lambda-i\rho)\epsilon_1} \; d\epsilon_1 = \frac{1}{\lambda -i\rho}
\end{equation}
Using contour integration around the pole $z=i \lambda$ (see also ref \cite{Integrals_book} p.318):
\begin{equation}\label{sigma(E)_3}
    \sigma(E) = \frac{\sqrt{N}}{2\pi} \lim_{\lambda \rightarrow 0} \int_{-\infty}^{+\infty} \frac{e^{-i\rho E}}{(\lambda - i\rho)^{N}} \; d\rho = \frac{\sqrt{N}}{(N-1)!} E^{N-1}
\end{equation}
On the other hand, the surface area $d\sigma(\epsilon)$ of the hyperzone of the hyperplane (Eq.(\ref{Heq})) where $\epsilon \leq \epsilon_1 \leq \epsilon +d\epsilon$ can be also calculated by extending the characteristic function of Eq.(\ref{sigma(E)dh}) as follows:
\begin{IEEEeqnarray}{rCl}\label{sigma(E)_4}
%a & = & b + c \\
d\sigma (\epsilon) & = & \sqrt{N} \cdot d\epsilon \int_{0}^{+\infty}\!\!\int_{0}^{+\infty} \cdots \int_{0}^{+\infty} \delta(\epsilon_1 + \epsilon_2 + \cdots + \epsilon_N - E) \delta(\epsilon_1 - \epsilon) \; d\epsilon_1 \; d\epsilon_2 \cdots d\epsilon_N
\nonumber\\
% & = & d + e \IEEEyesnumber
& = & \sqrt{N} \cdot d\epsilon \int_{0}^{+\infty}\!\!\int_{0}^{+\infty} \cdots \int_{0}^{+\infty} \delta(\epsilon_2 + \cdots + \epsilon_N - (E-\epsilon)) \; d\epsilon_2 \; d\epsilon_3 \cdots d\epsilon_N
\nonumber\\
& = & \sqrt{N} \cdot d\epsilon \int_{0}^{+\infty}\!\!
\int_{0}^{+\infty} \cdots \int_{0}^{+\infty} \Big\{ \frac{1}{2\pi}  \int_{-\infty}^{+\infty} e^{i \rho (\epsilon_2 + \epsilon_3 + \cdots + \epsilon_N - (E-\epsilon))} \cdot d\rho \Big\} \; d\epsilon_2 \; d\epsilon_3 \cdots d\epsilon_N \nonumber\\
& = & \frac{\sqrt{N}}{2\pi} \; d\epsilon \lim_{\lambda \rightarrow 0} \int_{-\infty}^{+\infty} e^{-i\rho(E-\epsilon)}\Big\{ \int_{0}^{+\infty} e^{-(\lambda - i\rho)\epsilon_2} \; d\epsilon_2 \Big\}^{N-1} d\rho \nonumber \\
& = & \frac{\sqrt{N}}{2\pi} \cdot d\epsilon \lim_{\lambda \rightarrow 0} \int_{-\infty}^{+\infty} \frac{e^{-i\rho(E-\epsilon)}}{(\lambda - i\rho)^{N-1}} d\rho = \frac{\sqrt{N}}{(N-2)!} (E-\epsilon)^{N-2} \; d\epsilon \IEEEyesnumber
\end{IEEEeqnarray}

From Eqs (\ref{sigma(E)_3}, \ref{sigma(E)_4}) we get the distribution of energy $P(\epsilon)$ of a particle of the system, according to the \textit{principle of equal a priori probabilities}:

\begin{equation}
    P(\epsilon) d\epsilon = \frac{d\sigma(\epsilon)}{\sigma(E)} \;\; \Rightarrow \;\; P(\epsilon)= \frac{N-1}{E} \Big( 1 - \frac{\epsilon}{E} \Big)^{N-2}
\end{equation}

At the limit $N \rightarrow \infty$; $E \rightarrow \infty$; $E/N = \langle \epsilon \rangle \sim$ fixed, we obtain the Boltzmann law:

\begin{equation}\label{P(e)final}
    P(\epsilon) = \lim \frac{N-1}{E} \big(1 - \frac{\epsilon}{E} \big)^{N-2} = \frac{1}{\langle \epsilon \rangle} e^{- \mathlarger{\frac{\epsilon}{\langle \epsilon \rangle}}}
\end{equation}

\section{Conclusion}\label{conclusion}

In the present work the statistics of a system containing $s$ quanta of equal energy $\epsilon_o$ distributed among $N$ particles was studied from two points of view.

\textbf{I.} By distributing the particles in the energy levels created by the quanta according to the Boltzmann principle of equal probability of configurations and the concept of average state, where the particles are distinguishable.\\
\textbf{II.} By distributing the quanta over the particles according to a principle of equal probability of states where the quanta are indistinguishable following the Bose statistics.

It was found that the two approaches lead to the same distribution $p(\kappa)$ (Eqs (\ref{probability_p(k)}, \ref{probability_p(k)_eq_k})) of the number of quanta contained in a particle. This shows that for the present system, the Bose statistics for the quanta is consistent with the Boltzmann principle for the classical particles.
%This change of point of view unites quantum and classical statistics. In particular, the latter quantum view that we have presented offers a major simplification of the two equations \ref{eq_particles_N_conserv} and \ref{eq_energy_conserv} into the single equation \ref{eq_k_cons_quantat}.
This change of point of view unites quantum and classical statistics for the present system. In addition, the latter quantum view that we have presented offers a major simplification in the statistical analysis of such systems by reducing the two Eqs (\ref{eq_particles_N_conserv}, \ref{eq_energy_conserv}) into the single Eq.(\ref{eq_k_cons_quantat}).
%It was additionally shown by an example that the concept of most probable configurations cannot be used for systems of finite $N$ and $s$.
This simplification becomes apparent when considering the geometrical representation of the two different cases as presented in the introduction.

At the classical limit $\epsilon_{o} \rightarrow 0$; $s \rightarrow \infty$; $s \epsilon_{o} = E \sim$ fixed, where the energies of the particles are continuous, both approaches give the Boltzmann law (Eq.(\ref{p(e)})) for $N \rightarrow \infty$; $E \rightarrow \infty$; $E/N =  \langle \epsilon \rangle \sim$ fixed. 
%At this limit, the transformation of the principle of equal probabilities of quantum states into the classical principle of equal a priori probabilities, valid on the $N$-th dimensional hyperplane (Eq.(\ref{Heq})), justifies the latter principle and gives a clear picture of the statistical foundation of the present problem.
This limit transforms the principle of equal probabilities of discrete quantum states studied here into the classical principle of equal a priori probabilities of continuous energies, valid on the $N$-th dimensional hyperplane (Eq.(\ref{Heq})), leading also to the Boltzmann's law (Eq.(\ref{P(e)final})). Thus the latter principle is justified by quantum mechanics providing a clear picture of the statistical foundation of the present problem.

\newpage
\pagenumbering{gobble}
\printbibliography

@BOOK {Boltzmann,
    author    = "L. Boltzmann",
    title     = "Lectures of Gas Theory",
    publisher = "University of California Press",
    year      = "1964"
}

@ARTICLE{DarwinFowler,
author = { C.G. Darwin and R.H. Fowler},
title = {On the partition of energy},
journal = {The London, Edinburgh, and Dublin Philosophical Magazine and Journal of Science},
pages = {450-479},
year  = {1922},
publisher = {Taylor & Francis}
}

@BOOK {Ladau,
    author    = "L. D. Landau and E. M. Lifshitz",
    title     = "Statistical Physics",
    publisher = "Pergamon Press",
    year      = "1968"
}

@BOOK {terHaar,
    author    = "D. ter Haar",
    title     = "Elements of Statistical Mechanics",
    publisher = "Butterworth-Heinemann",
    year      = "1995"
}

@BOOK {Integrals_book,
    author    = "I. S. Gradshteyn and I. M. Ryzhik",
    title     = "Table of Integrals, Series and Products",
    publisher = "Academic Press",
    year      = "1980"
}
\addcontentsline{toc}{part}{References}
\end{document}